\documentclass[useAMS,usenatbib]{mn2e}

\usepackage{graphicx}
\usepackage{multirow}
\usepackage{amssymb}

\title[Spectral Analysis of SN 2007od]{Quantitative Photospheric Spectral Analysis of the Type IIP Supernova 2007od}
\author[C. Inserra et al.]{C. Inserra$^{1}$$^{,}$$^{2}$$^{,}$$^{3}$\thanks{E-mail: cosimo.inserra@oact.inaf.it(CI)}, E. Baron$^{3,4,5}$,  M. Turatto$^{6}$\\
\\
$^{1}$Dipartimento di Fisica ed Astronomia, Universita' di Catania, Sezione Astrofisica, Via S.Sofia 78, 95123, Catania, Italy\\
$^{2}$INAF Osservatorio Astrofisico di Catania, Via S.Sofia 78, 95123, Catania, Italy\\
$^{3}$Homer L.~Dodge Department of Physics and Astronomy, University
of Oklahoma, Norman, OK 73019 USA\\
$^4$Hamburger Sternwarte, Gojenbergsweg 112, 21029 Hamburg, Germany\\
$^5$Computational Research Division, Lawrence Berkeley
        National Laboratory, MS 50F-1650, 1 Cyclotron Rd, Berkeley, CA
        94720 USA\\
$^{6}$INAF Osservatorio Astronomico di Trieste, Via Tiepolo 11, 34143, Trieste, Italy}

\def\kms{km\,s$^{-1}$}

\def\od{SN~2007od}

\begin{document}

\date{Received.....; accepted...........}

\pagerange{\pageref{firstpage}--\pageref{lastpage}} \pubyear{}

\maketitle

\label{firstpage}

\begin{abstract}
We compare and analyze a time series of spectral observations obtained during the first 30 days of evolution 
of SN 2007od with the non-LTE 
code \texttt{PHOENIX}.
  Despite some spectroscopic particularities in the Balmer features,
  this supernova appears to be 
  a normal Type
  II, 
  and the fits proposed are generally in good agreement with the observations. 
  As a starting point we have carried out an analysis with the parameterized synthetic
  spectrum code \texttt{SYNOW} to confirm line identifications
  and to highlight differences between the results of the two codes.
  The analysis 
  computed using \texttt{PHOENIX} suggests the presence of
  a high velocity feature in H$_{\beta}$ and an 
  H$_{\alpha}$ profile reproduced with a density profile steeper than that of
  the other elements.
  We also show a detailed analysis of the ions velocities of the 6 synthetic spectra. The distance is estimated for each epoch with the 
  Spectral-fitting Expanding Atmosphere Method (SEAM). Consistent results are
  found using all the spectra which give the explosion date of JD 2454403
  (29 October, 2007) and a distance modulus $\mu= 32.2\pm0.3$.
\end{abstract}

\begin{keywords}
  supernovae: general - supernovae: individual: SN 2007od - line:
  identification - galaxies: distances and redshifts
\end{keywords}

\section{Introduction}\label{sec:intro}

\od\/ was discovered on 2007 November 2.85 UT in the nearby galaxy UGC
12846 \citep{c1}. \citet{c2} classified it as a normal SN IIP about
two weeks after explosion, and reported some similarity with the
spectrum of the type II SN 1999em, 10 days after explosion. 
\od\/
exploded in the Magellanic Spiral (Sm:) galaxy UGC 12846, which has a
heliocentric recession velocity of 1734$\pm$3 \kms\/, a distance
modulus of 32.05$\pm$0.15 and an adopted reddening of
E$_{\mathrm{tot}}$(B-V)=0.038 \citep{07od}.  Extensive studies of the
photospheric and nebular periods have been presented in \citet{andrews}
and \citet{07od}.

\od\/ showed a peak and plateau magnitude, M$_{V}=-18.0$ and
M$_{V}=-17.7$ respectively \citep{07od}, brighter than common SNe IIP
\citep{p1,richardson}, but the luminosity on the tail is comparable
with that of 
the faint SN 2005cs \citep{05cs}. The luminosity on the tail was
affected by the early formation of dust \citep[$\lesssim$220d after
explosion,][]{andrews}. Based on mid infrared (MIR) observations in the nebular
phase, the amount of dust has been determined to be up to 4.2 x 10$^{-4}$
M$_{\odot}$ \citep[]{andrews} and the ejected mass $^{56}$Ni $\sim$
2$\times$10$^{-2}$ M$_{\odot}$\citep{07od}.


This object also shows interaction with a circumstellar medium
(CSM). There is  some evidence, in the form of high velocity features, for 
weak interaction soon after the outburst and solid observational
evidence for  interaction in the nebular phase \citep{07od}.  
However, due to the good temporal coverage, the position inside the host
  galaxy, and the low observed reddening, this supernova 
is a good candidate for analysis by the generalized stellar
  atmosphere code \texttt{PHOENIX} in order to learn more about its
physical structure.
In
\S~\ref{sec:model} we present the codes and the strategy
applied for modeling. In \S~\ref{sec:sp} we show the synthetic spectra
and the comparison with observed spectra, while in
\S~\ref{sec:ana} we provide an analysis of the principal
characteristics of the synthetic spectra.
A conclusion follows in \S~\ref{sec:final}.

\section{Method}\label{sec:model}

The preliminary line identification in the
spectra of \od\/, confirming those obtained by \citet{07od}, has been performed using
the fast, parameterized supernova synthetic spectra code \texttt{SYNOW}.
The code is discussed in detail by \citet{sy} and recent applications
include \citet{branchsy}, \citet{Ibcsy}, and \citet{roy}. \texttt{SYNOW}
assumes the Schuster-Schwarzschild approximation and the source
function is assumed to be given by  resonant scattering, treated in the
Sobolev approximation. It correctly accounts for the effects of
multiple scattering. 

For a subsequent, more detailed analysis,
  we have used the generalized stellar atmospheres 
code \texttt{PHOENIX} \citep{hb04,hb99}. 
The code includes a large
number of NLTE and LTE background spectral lines and solves
the radiative transfer equation with a full characteristics piecewise
parabolic method \citep{ha92}
without simple approximations such as the Sobolev approximation
\citep{mi70}.
The process that solves the radiative transfer and the rate equations with the condition 
of radiative equilibrium, is repeated until the radiation field and the matter
converge to radiative equilibrium in the Lagrangian frame.
These
calculations assume a compositionally homogeneous atmosphere with a power law
density 
and steady state conditions in the atmosphere.

\section{Spectra models}\label{sec:sp}

We have modeled the first six observed
photospheric spectra, covering a period from 5d to 27d since the
adopted explosion date JD= 2454404$\pm$5 \citep[$\sim30$ Oct. 2007,][]{07od}.  The most
interesting spectra are the first one of the series (5d), with a flat top
H$_{\alpha}$ profile and two uncommon features at about
4400\AA\/ and 6250\AA\/, and the last  one (27d), which has the
best signal to noise ratio among the plateau spectra. The detailed, comparative study
of these epochs provides important 
information about the presence of 
ions and  possible CSM
interaction at early times.

\begin{figure}
 \includegraphics[width=\columnwidth]{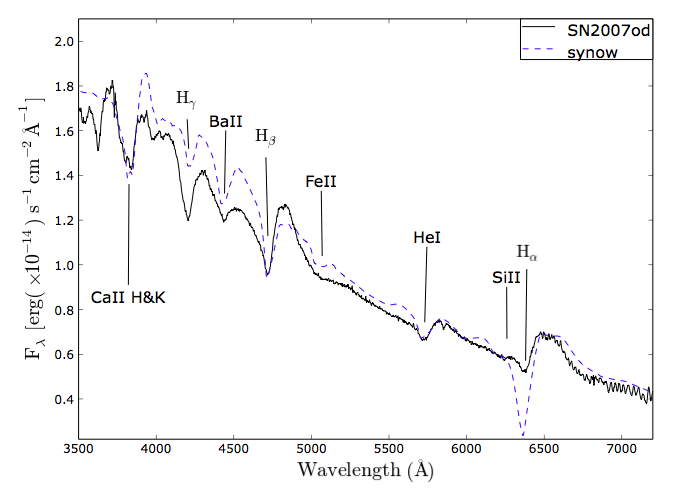}
 \caption{Comparison between the optical spectrum of \od\/ at 5 days post
   explosion (JD 2454404) and the \texttt{SYNOW} synthetic spectrum (for
   composition of synthetic spectra see text).}
 \label{fig:id1}
\end{figure}

\begin{figure}
 \includegraphics[width=\columnwidth]{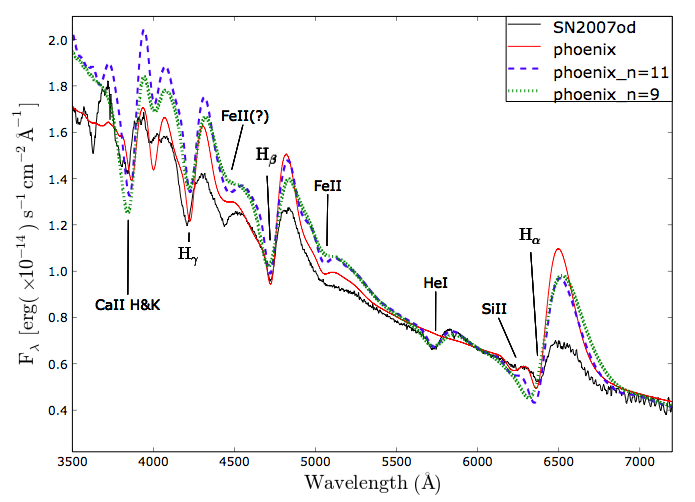}
 \caption{Comparison between the optical spectrum of \od\/ at 5 days post
   explosion (JD 2454404) and \texttt{PHOENIX} full NLTE spectra. The model
   parameters are those of the first row of Tab.~\ref{table:par} but for the n index reported in the legend.}
 \label{fig:id2}
\end{figure}

Figure~\ref{fig:id1} shows the line identifications determined by the
\texttt{SYNOW} analysis for the first spectrum,
obtained using 
T$_{\mathrm{bb}}\sim 12000$~K,  v$_{\mathrm{phot}}\sim 7800$~\kms\/,  optical depth
$\tau(v)$ parameterized as a power law of index $n =$ 9, and
T$_{\mathrm{exc}}$=10000 K assumed to be the same for all ions. The features
visible in Fig.~\ref{fig:id1} are produced by only 6 chemical
species. The P-Cygni profile of the Balmer lines are clearly visible, as well as
He~I $\lambda$5876, and significant contributions due to Ca~II, Fe~II,
Ba~II, and Si~II. The Balmer lines have been detached from the photosphere to
better match the observed velocity. The uncommon
lines mentioned above are identified as Ba~II ($\lambda$4524)
and Si~II ($\lambda$6355). In our attempts we have considered also the possible
presence on N~II $\lambda$4623, but the poor fit and the lack of
N~II $\lambda$5029 and N~II $\lambda$5679 lines (stronger than the first one), 
lead us to the conclusion that there is no 
enhanced N in the spectra of \od\/ \citep[cfr.][]{07od}.

\begin{figure*}
 \includegraphics[width=18cm]{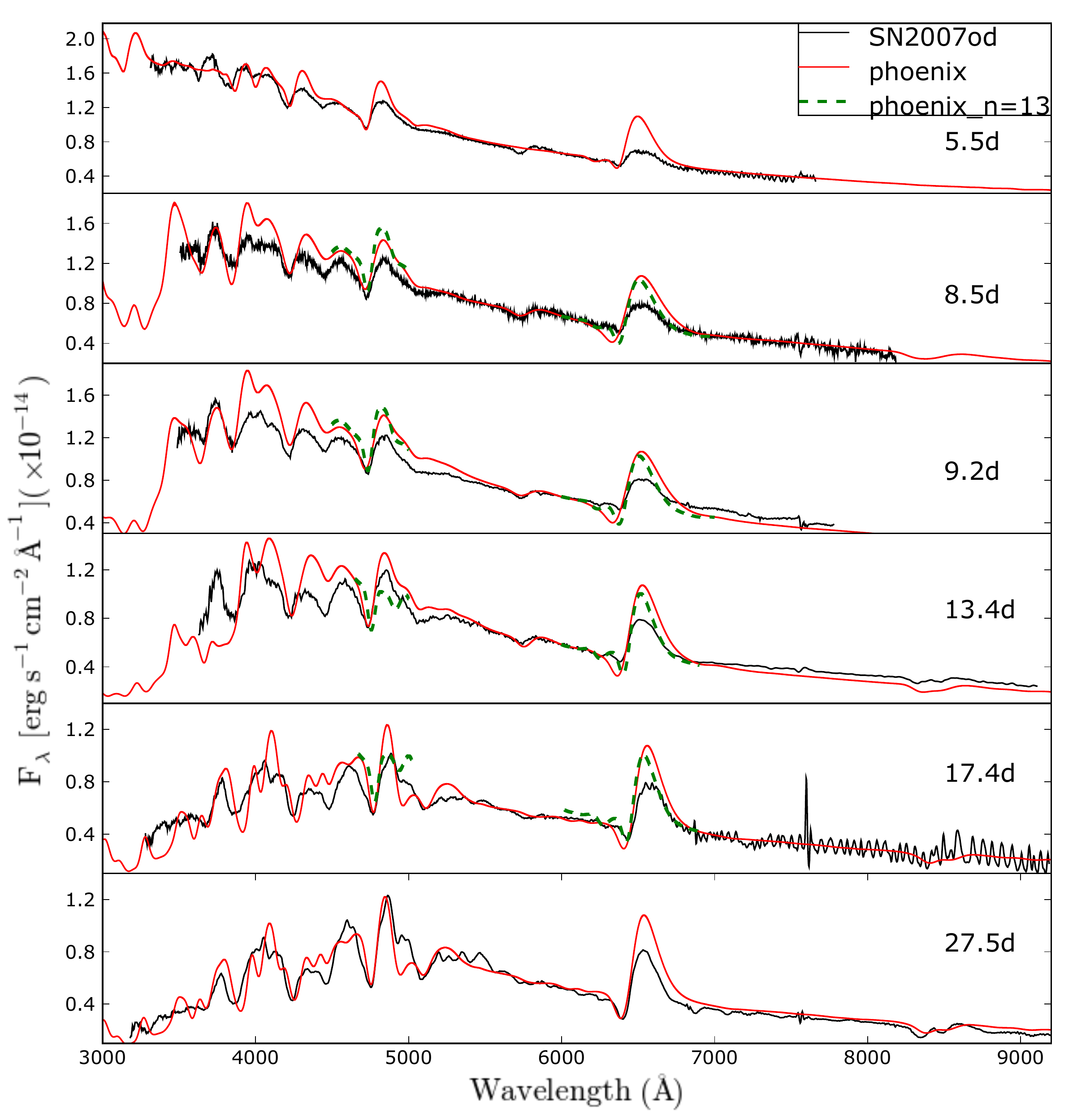}
 \caption{\texttt{PHOENIX} spectral evolution compared with observed spectrum. Model parameters for best fit spectra 
 (red) are listed in Tab.~\ref{table:par}. The dashed green lines in the spectra between day 8.5 and 17.4 
 in the H$_{\beta}$ and H$_{\alpha}$ regions are the line profiles computed with a density index n=13.} 
 \label{fig:ev}
\end{figure*}

With the adopted reddening (E(B-V)=0.038)
and the ions suggested by the \texttt{SYNOW} analysis, we have computed a grid of detailed
fully line-blanketed \texttt{PHOENIX} models.  We have explored variations in
multiple parameters for each epoch adjusting the total bolometric
luminosity in the observer's frame (parameterized by a model temperature,
T$_{\mathrm{model}}$), the photospheric velocity (v$_0$), the metallicity \citep[the solar abundances were those of][]{grevesse98} and the
density profile (described by a power law $\rho \propto r^{-n}$).
Gamma-ray deposition was assumed to follow the density profile.
We noticed an increase of the emission profiles, especially those of Balmer lines, 
with the increase of gamma ray deposition.
However, in our final models the gamma-ray deposition was {\bf not
included}.
We have estimated the best set of model parameters by performing a simultaneous $\chi^{2}$ fit of
the main observables.
The relevant parameters for \od\/ 
for the entire early evolution
are reported in Tab.~\ref{table:par} while the fits
are shown in Fig.~\ref{fig:ev}.

\begin{table}
  \caption{Parameters of \texttt{PHOENIX} models of \od.}
  \begin{center}\tabcolsep=0.5mm
  \begin{tabular}{cccccccc}
  \hline
  \hline
  JD & Phase$^{*}$ &T$_{model}^{\diamond}$ & v$_0$ & n$^{\dagger}$ & r & L\\
   +2400000 & (days) & (K) & (\kms)&& (10$^{14}$ cm)& (10$^{41}$ erg s$^{-1}$)\\
 \hline
  54409.5 & 5.5   &  8000 & 7600 &13 &3.6& 3.8\\
  54412.5 & 8.5    & 7400 & 7200 &9 &5.3& 6.0\\
  54413.2 &9.2     & 7300 & 7050 &9 &5.6& 6.3\\
  54417.4 &13.4   & 6800 & 6000 &9 &6.9 & 7.2\\
  54421.4 &17.4   &  6200 & 5400 &9&8.1& 6.9\\
   54431.5 &27.5   &  6000 & 5000 &9&11.9& 13.1\\
 \hline
\end{tabular}
\begin{flushleft} 
$^{*}$ with respect to the explosion epoch (JD 2454404) from \citep{07od}\\
$^{\diamond}$ with a total E(B-V)=0.038 \citep{07od}\\
$^{\dagger}$ index of power law density function
\end{flushleft} 
\end{center}
\label{table:par}
\end{table}

\begin{figure}
 \includegraphics[width=\columnwidth]{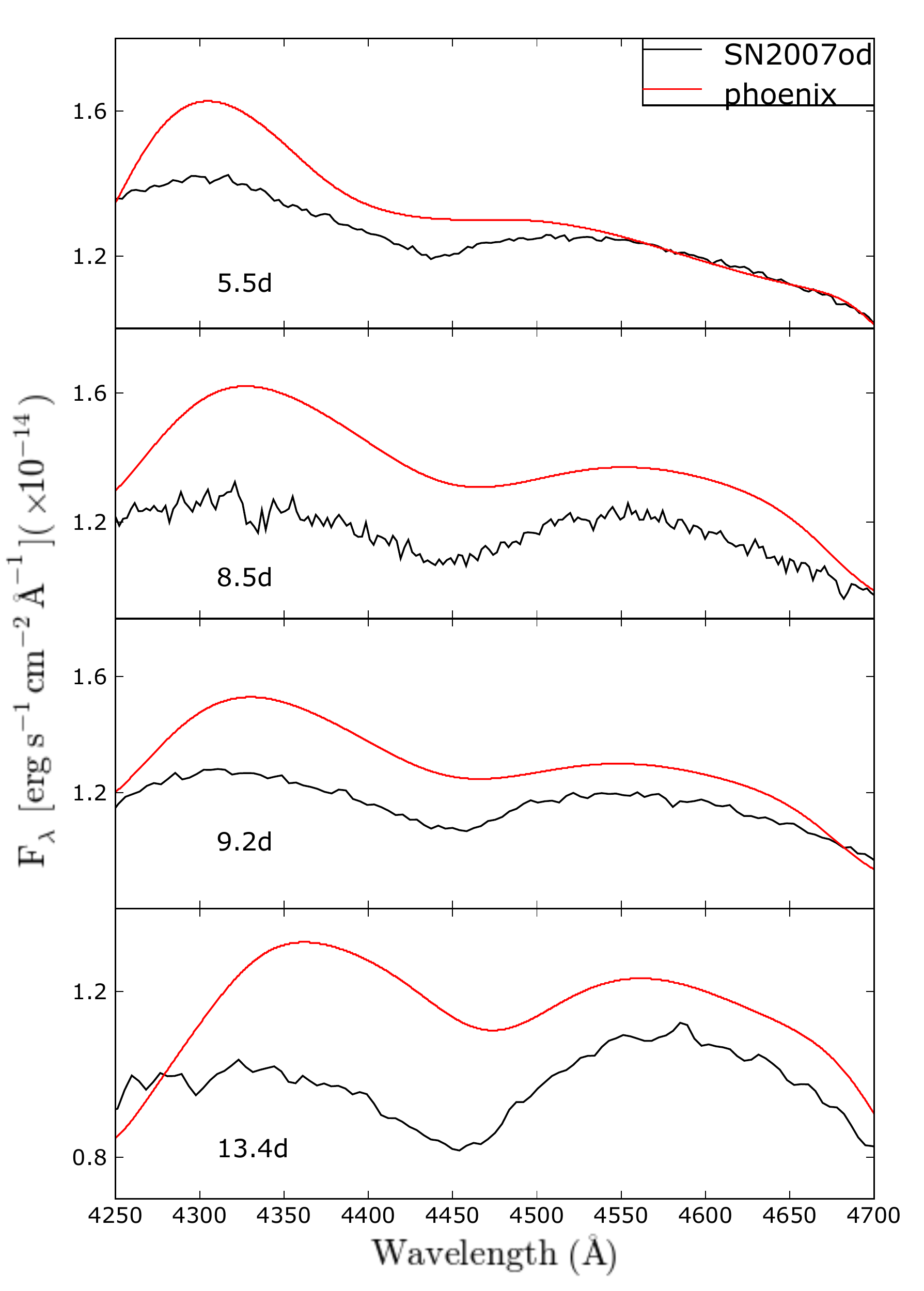}
 \caption{A blowup of the 4440\AA\/ region where the
   \texttt{PHOENIX} synthetic spectra are  compared to observed {\bf ones}.}
 \label{fig:z}
\end{figure}

In the first \texttt{PHOENIX} spectra we have treated in NLTE the following
ions:
H~I, He~I-II, Si~I-II, Ca II, Fe~I-II, Ba~I-II. 
He~I, Si~I, Fe~I and
Ba~I have been considered to reproduce the ionization levels of the
corresponding atoms. The opacity for all other ions is
treated in LTE with a constant thermalization parameter $\epsilon =
0.05$ \citep[see][for further details]{baron96b}. As shown in Fig.~\ref{fig:id2} there is no line that
corresponds to 
Ba~II, even though the
identification for the absorption at 4400\AA\/ seemed plausible in the
\texttt{SYNOW} analysis. The temperature is too high 
to produce a Ba~II line, all the more so with the observed
strength.
The presence of Ba~II was checked by calculating a set of single ion spectra, that
is calculating the spectrum with all continuum opacities, and only
lines from Ba~II as well as via the inverse procedure of turning off
the line opacity from Ba~II. The same procedure was performed
for the He~I ion, in order to study the feature that could arise around 4471\AA\/ but no significant
contribution can be seen in the synthetic spectra. 
The closest line to the 4440\AA\/ feature is due to Fe~II, 
though it is not as strong as in the observed spectrum. The
evolution of the 4440\AA\/ region (shown in Fig.~\ref{fig:z}) displays
the inconsistency.  The Fe~II line explains the feature starting from
day 9, though the velocity does not match the observed line position.  
It is likely that the observed profile of the 4440\AA\/ feature
is due to the combination of this line with high velocity feature (HV)
of $H_\beta$, formed by an increased line opacity that our simplified model
($\rho\propto r^{-n}$) is not able to reproduce \citep[the HV features are discussed in][]{07od}. 
Though Ba~II identification provided by \texttt{SYNOW} is not
completely ruled out by the NLTE analysis, however we consider it unlikely.


The presence of Si~II at 6530\AA\/ is confirmed by the \texttt{PHOENIX}
analysis, despite the weakness of the synthetic feature. In Fig.~\ref{fig:id2}
it is barely visible, but the same analysis performed for Ba~II
confirms its 
identification.
The presence of Si~II in SNe IIP is not uncommon. The other absorption
features are successfully reproduced, except for H$_{\alpha}$ (see
\S~\ref{sec:ana}) and the absorption lines in the region of
3600\AA\/ that are possibly related to Ti~II, not included in
the first NLTE spectra in order to minimize CPU time. 
As shown in Fig.~\ref{fig:id2} we tried different density indices to better match the entire profile. 
The best match for the overall spectrum is given by the model with n=13 (in all figures the best 
 \texttt{PHOENIX} model is always plotted in red), even if the strengths, with respect to the normalized continuum, of H$_{\gamma}$ 
 and
 He~I are better reproduced by the models with a density exponent lower than $n=13$.
 The steeper density profile leads to greater emission than models with flatter density profiles.

\begin{figure*}
 \includegraphics[width=18cm]{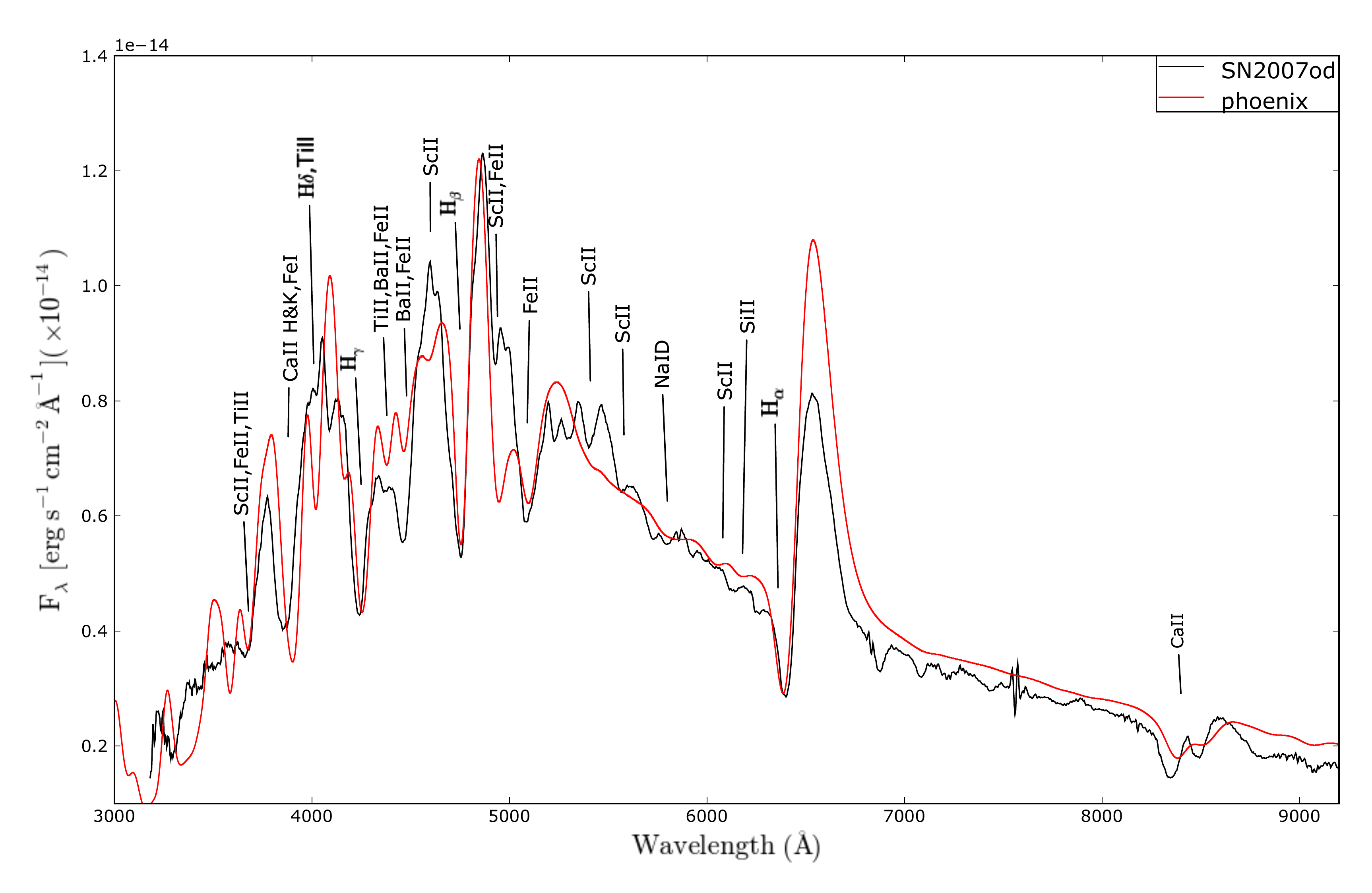}
 \caption{Comparison between the optical spectrum of \od\/ at 27 days post
   explosion (JD 2454404) and the \texttt{PHOENIX} full NLTE spectrum (for model
   parameters see Tab.~\ref{table:par}).}
 \label{fig:id3}
\end{figure*}

Both the second (8.5d) and
third (9.2d) spectra have been constructed using the same setup that
was used to calculate the
first spectrum, except that the density index was decreased from 13 to 9.
A profile steeper than n=9,  better reproduces the absorption profile of the Balmer
lines, as seen in first epoch, but all the models show
flux  slightly higher blueward of  5500\AA\/ and especially at about 4000\AA\/. 
Also the strength of H$_{\gamma}$ is greater than observed in the model with $n>9$.
The flux on the blue side of Ca II H\&K is lower than observed. 
 We found that a steeper density profile enhances  the discrepancy
 between the observed and synthetic H\&K line profile.

 In general, the line profiles suggest a  density distribution
 more complex that a single power law. Our calculations suggest that
 perhaps a broken power law would better reproduce the observed
 spectra, with a shallower density profile at lower velocities and a
 steeper profile at higher velocities.
 The Ca~II or He~I ions are clearly better reproduced by a density index close to 9,
 while the metal elements (e.g. Fe and Si) which form close to the photosphere (and are
 relatively weak) are less sensitive to the assumed density profile.
The Fe~I-II and Si~II are well reproduced with all density profiles, even if the steeper profile better reproduces the Si~II profile than the flatter profile (see Fig.~\ref{fig:id2}).
While the more complex density profile could be intrinsic in the
initial structure it is also possible that 
early interaction of the SN ejecta with a close in circumstellar region affect the
line profiles. Indeed, the flat topped nature of the Balmer lines may
indicate circumstellar interaction. A broken power law density distribution of the ejecta has been claimed also by \citet{utchug11}
 in the case of SN 2000cb. To illustrate the case we have over-plotted in Fig.~\ref{fig:ev} the H$_{\beta}$ and H$_{\alpha}$
 regions obtained by n=13 models (green dashed line) for the spectra from 8.5d to 17.4d.

\begin{figure}
 \includegraphics[width=\columnwidth]{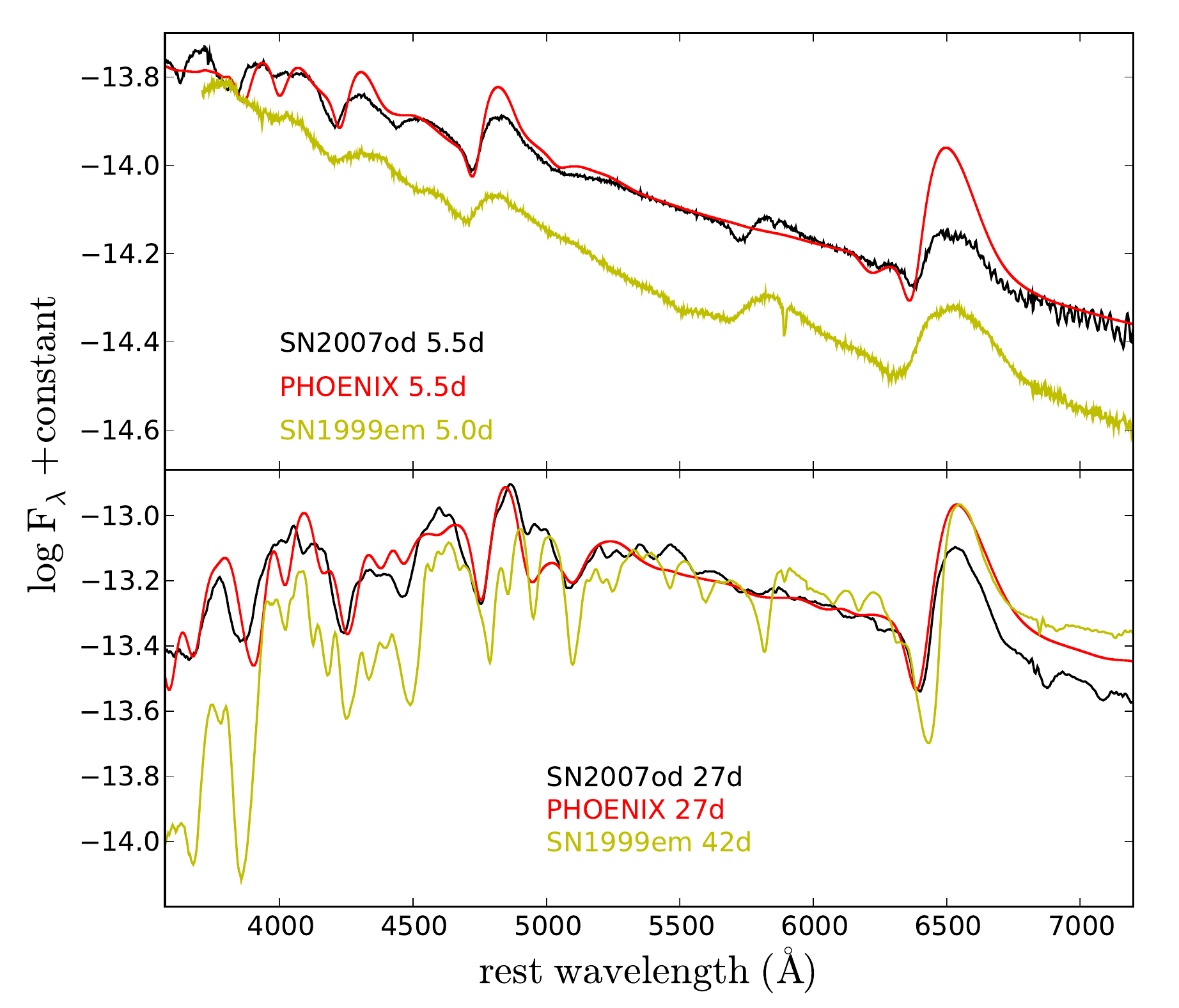}
 \caption{Comparison between observed spectra of \od\/, \texttt{PHOENIX} full NLTE spectra (for model
   parameters see Tab.~\ref{table:par}) and SN 1999em spectra at the first (top) and last epoch (bottom) of our series.}
 \label{fig:comp}
\end{figure}

The spectra at 13.4 and 17.4 days show the increasing effects of a few
metal lines such as Fe~II $\lambda$5169 and Sc~II $\lambda$6300 that
indicate the lower temperature at the beginning of the
plateau phase. Moreover, the presence of metal ions changes the flux
at $\sim$3800\AA\/ in the spectra at 13d.  From this epoch onward, the
4440\AA\/ feature seems more clearly related to Fe~II, strengthening our
conclusion for the absence of Ba II.
  
The final spectrum we analyze was obtained on  November 27, 2007
(27d) when the supernova is solidly on the plateau.
At this epoch the metal lines are fully developed, as is the Na~ID feature
that has replaced the He~I $\lambda 5876$ line. Thanks to
the broad wavelength coverage, the good resolution (11\AA\/), and the
high signal to noise (S/N=60), this is the best available spectrum of SN 2007od.

Fig.~\ref{fig:id3} displays our best NLTE model. The species treated
in NLTE are H~I, Na~I, Si~I-II, Ca~II, Sc~I-II, Ti~I-II, Fe~I-II and
Ba~I-II. Here, more than at earlier epochs, the
presence of the neutral species change the strength of some lines,
especially in the blue  ($<$ 4000\AA\/) and around 5000\AA\/. The
H lines, Na ID,  Fe~II $\lambda$5169,  Sc~II $\lambda$5658,
and $\lambda$6245 are well fitted. The Ca~II lines 
also are well fitted.  Other lines fit fairly well in terms of velocity width, but
not necessarily in total flux. This is due to the fact that many lines
are blended with others. 

All models have solar abundances and metallicity. We studied the
effects of reducing the model metallicity to that deduced for the \od\/ environment Z$<$0.004
\citep[see][]{07od}, but the changes
in the spectra were negligible at the 3$\sigma$
  level. 

The comparative evolution of SN 2007od has been discussed in Sect. 3.3 of \citet{07od}.
The most interesting features noticed in the earliest phases were the 4400\AA\/ absorption and the boxy profile of the Balmer lines.
In Fig.~\ref{fig:comp} we compare the spectra of SN 2007od at two phases (day 5.5 and day 27) with those of SN 1999em, the most similar SN IIP as determined by GELATO analysis \citep{avik}, and with the two corresponding best synthetic spectra.
Indeed the overall similarity is remarkable. However there are also interesting differences, in addition to the two major ones mentioned above. At the first epoch the expansion velocity is larger in SN 1999em (FWHM(H$_{\alpha}$)$_{99em}$$\sim$11000 \kms $>$ FWHM(H$_{\alpha}$)$_{07od}$$\sim$9000 \kms) while at the second epoch SN 2007od is faster then SN 1999em (FWHM(H$_{\alpha}$)$_{07od}$$\sim$7000 \kms $>$ FWHM(H$_{\alpha}$)$_{99em}$$\sim$5500 \kms).
However, the epochs are offset so evolution could play a role.
The slower velocity evolution of SN 2007od was already noticed by \citet{07od}. The model is able to reproduce consistently the spectra of SN 2007od at both epochs. The {\bf other} major difference is in the line contrast.
Considering the ongoing early interaction, we can interpret the effect as due to toplighting \citep{top} which smooths the line contrast.
At the second epoch the difference between the line profiles of the two objects and the model is reduced. 

\section{Analysis}\label{sec:ana}

\begin{figure*}
 \includegraphics[width=18cm]{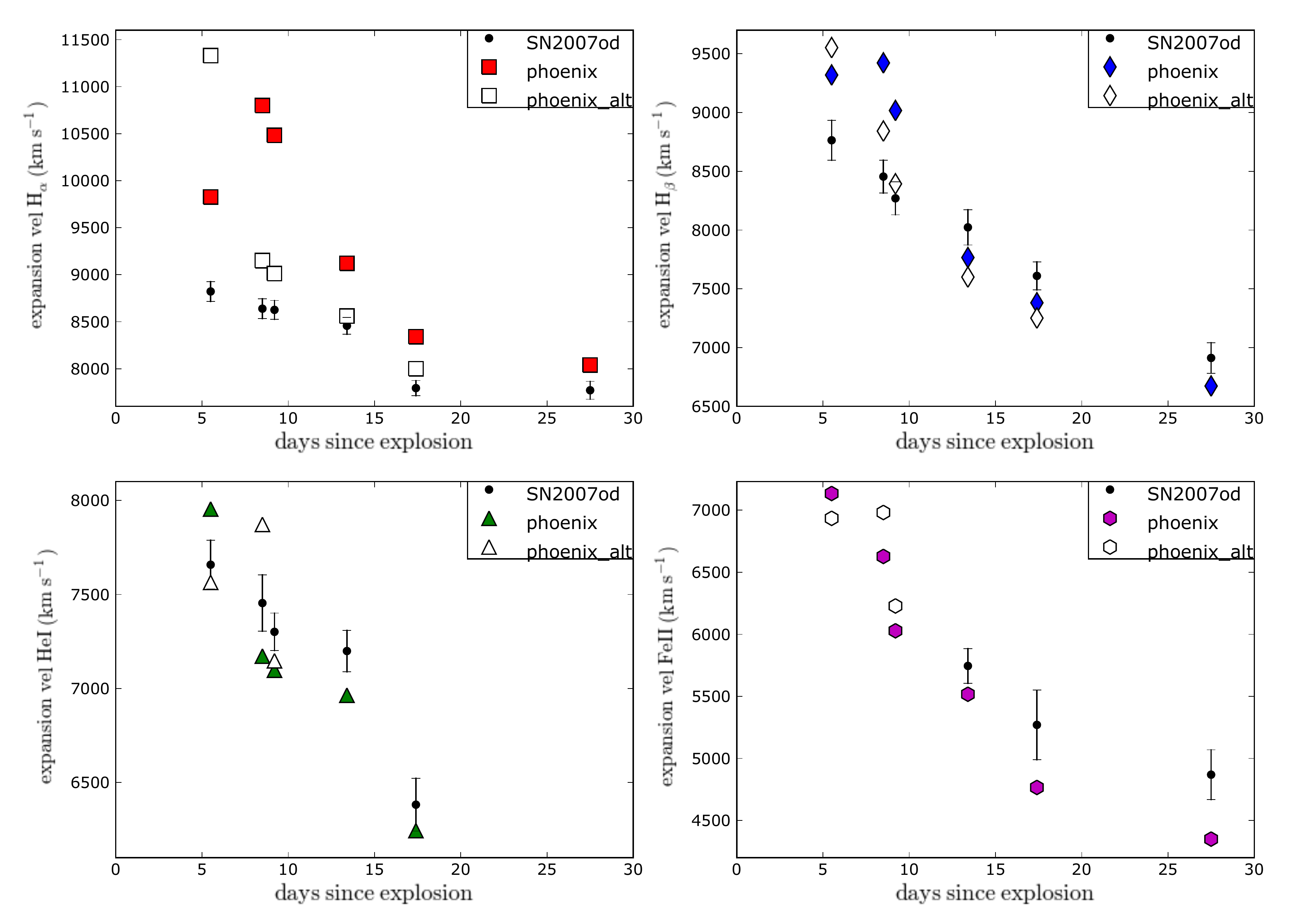}
 \caption{Expansion velocities of H$_{\alpha}$, H$_{\beta}$,
   He~I, and Fe~II measured from synthetic \texttt{PHOENIX} NLTE spectra compared with
   those measured from the observed spectra (black dots with errorbars) by \citet{07od}. Filled symbols refer to the spectra shown in Fig.~\ref{fig:ev}, while
open symbols to spectra calculated with different density indexes
($n=9$ for the first epoch, $n=13$ for the following).}  
 \label{fig:vel}
\end{figure*}

The expansion velocities of H$_{\alpha}$, H$_{\beta}$, He~I $\lambda 5876$,
and Fe~II $\lambda 5169$ as derived from fitting the absorption minima of
the \texttt{PHOENIX} spectra, are shown
in Fig.~\ref{fig:vel} together with the measured values in the observed spectra. 
The filled symbols indicate the spectra shown in Fig.~\ref{fig:ev} (reported in Tab~\ref{table:par}), while
the open symbols refer to spectra calculated with different density indexes
($n=9$ for the first epoch, $n=13$ for the following epochs).

In Fig.~\ref{fig:vel} (top left panel) and
Fig.~\ref{fig:ev} the absorption minima in the NLTE (n=9) spectra
for H$_{\alpha}$ are too blue when compared with the observed spectra,
particularly for the first three epochs. Instead the models with n=13 better reproduce
the observed velocity. This could be due to several
effects. The simple uniform power-law density
profile, assumed here, may be not accurate enough to describe 
the ejecta. Also ionization effects,
perhaps due to circumstellar interaction, may reduce the Balmer
occupations over those that are predicted in the $n=9$ models. Combined with
the evidence of the flat-top emission, it is somewhat  possible that
circumstellar interaction effects are responsible for the observed
shape of the Balmer lines and for the different density profiles.
The effect disappears with time. Also for
H$_{\beta}$, the \texttt{PHOENIX} and the observed velocity from 13.4d are
comparable within the errors for both models. From day 8.5 and onward the $n=9$ models better reproduce 
the H$_{\beta}$ profile in velocity and strength.
Neither He~I nor Fe~II show this
behavior in the early epochs, indicating that the effect is most
likely confined to the outermost layers of the ejecta. Except for the
epochs reported above, all the line velocities are slightly smaller
than observed. Though we have not included time-dependent rates in this
calculation, this effect seems not so likely to explain the
discrepancy between the velocity of the synthetic and observed
H$_\alpha$ \citep{utchug05,dessart07a,soma09,soma10}.

 Clearly our simplistic models do not reproduce the HV feature of H$_{\beta}$ and overestimate the emission
strength of H$_{\alpha}$, especially in the first spectrum.
While interaction of the SN ejecta with circumstellar matter may be important, it is also possible that the density distribution we have adopted is too simplistic. The latter highlights the need for using the correct physical structure when modeling the SNe.

The sample of photospheric NLTE spectra collected for the \od\/ 
{\bf allow to apply the}
Spectral-fitting Expanding Atmosphere Method
\citep[SEAM,][]{baron93,baron94,baron95,baron96} 
since in addition to a good overall shape of the spectra, the models predict
consistent fluxes. We can derive a
distance modulus by subtracting our calculated absolute magnitudes from
the published photometry \citep{07od}, assuming a reddening of E(B-V)$_{tot}$=0.038.
Considering all the epochs, we obtain
the distance modulus of the supernova, $\mu$=32.5$\pm$0.3
where the errors include the standard deviation of our fits
  and the error due to the uncertainty in the interstellar reddening.
Since SEAM is strongly dependent on the uncertainties of the
  explosion date, it seems reasonable
to give lower weight to early spectra, since a longer time baseline minimizes the error due to the uncertainties
in the explosion day.
If we consider only 
the later observed spectra, from 13.4d to
27.5d, we find
$\mu$=32.2$\pm$0.2, in good agreement with the value reported in
\citet{07od}. 
Furthermore, model spectra for the last three epochs better fit the observations over the entire spectral range, and the derived distance modulus is comparable with the \citet{mo00} measurement within the uncertainties. 

Indeed, 
a good agreement with the \citet{07od} distance ($\mu$=32.2$\pm$0.3)
is obtained by using all spectra and
by taking the explosion date to be JD 2454403 (29 October, 2007).
Hence we conclude that the
explosion date is October 29 $\pm$1.5 days. The explosion
date should always be determined by a $\chi^2$ minimization in a SEAM analysis.


\section{Conclusions}\label{sec:final}

We have shown that both direct synthetic spectral fits and detailed
NLTE models do a good job in reproducing the observed optical spectra
of \od\/. We have also pointed out that detailed NLTE spectral modeling of early
spectra does not support the line identification (Ba~II, Fe~II or He~I) of the
4440\AA\/ as suggested by the \texttt{SYNOW} model or as identified in
previous SNe~IIP.  Rather we interpret the line as a combination of
Fe~II and with an additional contribution from a possible HV feature of
H$_{\beta}$.  The origin of the line could be due to 
increased opacity at high velocities due to variation in the level
populations not accounted in the simple density profile of
\texttt{PHOENIX} or due to possible interaction in the outermost
layers. This result suggests extreme caution with line 
identifications in hot, differentially expanding flows. At the
same time it shows that \texttt{PHOENIX} calculations give reliable
results and
lend support to the reliability of the modeling.  Our spectral
analysis shows that during the plateau Sc~II lines can arise even with
standard solar abundances.  Also Si~II lines are reliably identified
in the spectra of our relatively normal SN IIP at early times. Another important issue is related to 
the difference between the density profile ($\rho\propto r^{-13}$)
needed to reproduce  H$_\alpha$ from 8.5d to 17.4d and the density law
required to 
reproduce the other elements ($\rho\propto r^{-9}$).
The evidence for a broken power law density profile shown by SN 2007od could be a general property of type II SNe
or maybe a consequence of a perturbed ejecta, especially in the outermost layers. 
We have also {\bf discussed} the velocity and
strength evolution of the principal lines. We have also applied the
Spectral-fitting Expanding Atmosphere Method (SEAM) to \od\/ obtaining  good
agreement with the distance modulus provided in \citet{07od} which allows to better constrain 
the explosion date (29 October, 2007).


\section*{Acknowledgments}

This work was supported in part NSF grant AST-0707704, and US DOE
Grant DE-FG02-07ER41517 and NASA program number HST-GO-12298.05-A.
Support for Program number HST-GO-12298.05-A was provided by NASA
through a grant from the Space Telescope Science Institute, which is
operated by the Association of Universities for Research in Astronomy,
Incorporated, under NASA contract NAS5-26555.  C.I. thanks David
Branch for the useful discussions.  M.T. is partially supported by the
PRIN-INAF 2009 \textquotedblleft Supernovae Variety and Nucleosynthesis Yields".
We thank the anonymous referee for the useful suggestions that improved our paper.

\label{lastpage}

\end{document}